\documentclass{kerauth}
\usepackage{graphicx}
\usepackage[round]{natbib}
\usepackage{amssymb}
\usepackage{epstopdf}
\usepackage{mathtools}
\usepackage{url}
\usepackage{floatrow}
\usepackage{pdflscape}
\usepackage{tabularx}
\begin{document}
\KER{1}{24}{00}{0}{2017}{S000000000000000}

\runningheads{Ewa Andrejczuk et al.}{The composition and formation of Effective Teams. Computer Science meets Organisational Psychology}


\title{The Composition and Formation of Effective Teams. Computer Science meets Psychology}

\author{Ewa Andrejczuk\affilnum{1,3},
Rita Berger\affilnum{2},
Juan A. Rodriguez-Aguilar\affilnum{1},
Carles Sierra\affilnum{1} and
V\'{\i}ctor Mar\'{\i}n-Puchades\affilnum{2}}

\address{\affilnum{1}IIIA-CSIC, Campus UAB, 08193 Bellaterra, Catalonia, Spain\\
\email{\{ewa,jar,sierra\}@iiia.csic.es,}\\
\affilnum{2}University of Barcelona, Barcelona, Spain\\
\email{ritaberger@ub.edu, vmarinpu7@alumnes.ub.edu}\\
\affilnum{3}Change Management Tool S.L., Barcelona, Spain\\}

\begin{abstract}
Nowadays the composition and formation of effective teams is highly important for both companies to assure their competitiveness and for a wide range of emerging applications exploiting multiagent collaboration (e.g. crowdsourcing, human-agent collaborations). The aim of this article is to provide an integrative perspective on team composition, team formation and their relationship with team performance.
Thus, we review the contributions in both the computer science literature and the organisational psychology literature dealing with these topics. Our purpose is twofold. First, we aim at identifying the strengths and weaknesses of the contributions made by these two diverse bodies of research. Second, we pursue to identify cross-fertilisation opportunities that help both disciplines benefit from one another. Given the volume of existing literature, our review is not intended to be exhaustive. Instead, we have preferred to focus on 
the most significant contributions in both fields together with recent contributions that break new ground to spur innovative research.
\end{abstract}

\section{Introduction}
The latter part of the 20\textsuperscript{th} and the beginning of the 21\textsuperscript{st} centuries have  witnessed a significant transformation from work organized around individual jobs to team-based work structures together with a focus on organisational efficiency \citep{kozlowski}. This is due to the increasing complexity of tasks, which in many cases cannot be performed by single individuals \citep{Ramezan}. Additionally, changes in technology facilitate workers in distinct locations to communicate and collaborate at low or no cost. On that account, team composition and formation research is of interest to many fields of science, primarily to organisational psychology. Moreover, it has also substantially pervaded the field of computer science, mainly within the area of multiagent systems (MAS). Indeed, research in MAS has considered a variety of application domains (e.g. Unmanned Aerial Vehicle (UAV) operations \citep{haque2013multilevel}, teamwork in social networks \citep{Lappas:2009:FTE} or RoboCup rescue teams \citep{Ramchurn:2010:DCR}) wherein agents face the challenge of performing tasks that are either too complex for one single agent or limited in time, thus requiring several agents to collaborate. 


Nevertheless, research on team composition and team formation in computer science (CS) and organisational psychology (OP) has evolved separately. On the one hand, MAS literature has typically disregarded significant OP findings, with the exception of several recent, preliminary attempts (such as \cite{FarhangianPPS15}, \cite{Hanna2015}, \cite{Andrejczuk}). Thus, this body of research has focused on algorithms that help automate team formation and composition.
On the other hand, the OP literature has mainly focused on empirically investigating the factors that influence team performance to develop heuristics that help organisations handcraft their teams. OP has disregarded the algorithmic results developed by computer scientists to automate team composition and formation. Despite the common research interests shared by MAS and OP, to the best of our knowledge there has been no effort in the literature to bridge the knowledge produced by both research disciplines. 

Against this background, the aim of this article is to survey both disciplines, to analyse and compare the strengths and weaknesses of their contributions, and to identify research gaps and opportunities by bringing together the knowledge of the two research strands on team composition and formation. Our analysis also pursues to identify cross-fertilisation opportunities that help both disciplines benefit from one another.


In order to structure our analysis, we have identified several dimensions that help us dissect the contributions from both research fields:

\begin{enumerate}
\item \emph{WHO is concerned?} The properties of the agents involved.
\item \emph{WHAT is the problem?} The features of the task to complete by a team.
\item \emph{WHY do we do it?} The objective function to optimise when composing/forming a team.
\item \emph{HOW do we do it?} The organisation and/or coordination structure adopted by the team in charge of performing a particular task.
\item \emph{WHEN do we do it?} The dynamics of the stream of tasks to be completed by agent teams.
\item \emph{WHERE do we do it?} The context wherein team composition/formation occurs.
\end{enumerate}


Our analysis of the literature indicates that Computer Science (CS) and Organisational Psychology (OP) exhibit some similarities. 
Indeed, one of the crucial findings in both OP and CS is that team members have to be heterogeneous to maximize team performance. When modeling agents, CS and OP agree on considering two main approaches: either there is complete information about the properties of each agent; or agents are capable of learning about their teammates through repeated interactions. Regarding tasks, both OP and CS research largely focus on finding team members whose properties make them capable of performing a given task based on its requirements. In other words, they are both concerned with matching agents (or whole teams) with tasks.  

However, there are important differences between the contributions made by OP and CS that stem from the fact that OP does consider the whole complexity of: humans as team members, tasks, the \emph{context} where teams perform tasks (understood as the internal and external factors influencing teamwork), and the dynamics of the actual-world scenarios where tasks appear to be serviced. Thus, OP assumes that human capabilities are necessarily dynamic (evolve along time) so that teams can successfully perform tasks in dynamic real-world scenarios and in a variety of contexts. Furthermore, OP observes that the quality of human resources (e.g. motivation, satisfaction, commitment), the ability of individuals to learn new capabilities, and the context constraining team performance significantly influence team performance. Finally, OP research also focused on identifying correlations between task types and team types to compose the best team depending on the type of each particular task. All these findings contributed by OP research offer interesting opportunities for cross-fertilisation.


The rest of the paper is organised as follows. Section \ref{Back} introduces some fundamental terminology to make clear what we mean by team composition, team formation and teamwork. Thereafter, the paper is organized around two main sections.  Section \ref{CS} reviews the MAS contributions to team composition and team formation. Next, section \ref{OP} surveys the contributions in the organisational psychology literature. 
Rangapuram2015Finally, section \ref{Dis} identifies the main similarities and differences between the two bodies of research. Furthermore, it also discusses cross-fertilisation opportunities between both fields that may spur future research.

\section{Background}\label{Back}

We introduce the fundamental terminology used in this survey. We refer to:
\begin{enumerate}
\item Team Composition as the process of deciding which agents will be part of a team,
\item Team Formation as the process of learning by agents to work together in a team and through this learning decide the roles and internal organisation of a team,
\item Teamwork as the process of performing a task by a composed and formed team.
\end{enumerate}

While there is a common understanding of teamwork within both OP and CS, the scientists do not agree on the notion of team formation. In computer science it is mostly understood as the process of deciding which agents will be a part of a team (here called team composition). Our definition of team formation is in line with the organisational psychology literature \citep[p.16]{kozlowski}. 

Another discrepancy between the computer science and the organisational psychology literature is the notion of skill and competence. Typically in computer science all kinds of agents' competences are called skills, while in OP the definition is more complex. In OP a prominent conceptualization of competence was given by Roe \citep[p.195]{roe2002}. He defines competence as ``a learned ability to adequately perform a task, duty or role''. Following his definition competences ``integrate knowledge, skills, personal values, and attitudes and are build on knowledge and skills and are acquired through work experience and learning by doing'' \citep{bartram2005}. 
Hence, competences include abilities and behaviours, as well as knowledge that is fundamental to the use of a skill. An example may consist of a programming task.  In order to effectively write a script one needs good logical and analytical competences as well as the skill to write a program in a specific language. Hence, Java is a skill.  Although, underlying the ability to use that skill effectively is a competence. 

\section{Team composition and formation from a computer science perspective}\label{CS}

Team composition and formation are critical issues for co-operative multiagent systems. In this section we survey the most recent and representative approaches in the MAS literature to the team composition and formation problems along the dimensions identified in the introduction above.



\subsection{WHO is concerned?}\label{skills}

The question behind team composition and formation is how to create a multiagent system as a group of heterogeneous agents (such as humans, robots, software agents or even animals) and how to organize their activities. Team members must observe the environment and interact with one another in order to perform tasks or solve problems that are beyond their individual capabilities. The algorithms to create these teams take inspiration from human teamwork. We observe people working together on daily activities as well as on research and business projects. For instance, there are  sport teams (e.g. football, basketball), police squads, search and rescue teams formed by dogs and humans, and we start to witness human-robot cooperation in houses, hospitals, or even in space missions \citep{Hoffman}.

In general, MAS research focuses on the interaction among intelligent agents. In the team formation literature, the focus is on the interaction of cooperative and heterogeneous agents. That is, agents who share a common goal, and have different individual properties. Therefore, in this section, we would like to account for the different ways previous research has dealt with these questions. We will classify individual properties according to two dimensions:
\begin{enumerate}
\item Capacity: individual and social capabilities of agents; and
\item Personality: individual behaviour models.
\end{enumerate}

\subsubsection{Capacity: individual and social capabilities of agents}\label{capabilities}  In many domains, a capability is defined as a particular skill required to perform an action. The capacity dimension has been exploited by numerous previous works, like Robust Team Formation \citep{Crawford,Okimoto} or Online Team Formation \citep{Anagnostopoulos12onlineteam}. In these works, agents are assumed to have multiple binary skills (i.e., the agent either has a required skill or not). This is a simplistic way to model an agent's capabilities since it ignores any skill degree. In real life, capabilities are not binary since every individual (e.g. human or robot) shows different action performance. 
This is why some works propose a more realistic approach by defining graded agent capabilities, for instance by defining skill levels \citep{Chalkiadakis2012}. 

On a different vein, \cite{Rangapuram2015} builds a weighted, undirected graph where the weight between each pair of agents reflects their degree of compatibility to jointly solve tasks. These weights are updated along multiple encounters between agents. In a somehow related vein, \cite{JAR2015} try to capture the quality of the solutions of team tasks via a model that besides using skills and compatibility between agents (called the strength of collaboration synergies within coalitions), calculates the reputation of teams (coalitions) as a whole and of single agents. These reputation values are used by the team composition process.

Typically, the capabilities of agents are assumed to be known, though there exist models that consider that an agent can learn the capability levels of other agents. For instance, \cite{Liemhetcharat2014} had the insight that repeated interactions allow to discover the capabilities of other agents. They call ``synergy'' to the degree of performance of a team. Agents learn a model of synergy via repeated interactions. Such synergy values are then used by individual agents to learn the capabilities of others, and hence to subsequently compose teams with improved performance. 
However, in open environments (that is, when new agents and tasks are dynamically introduced), agents need more sophisticated procedures to decide which team to join. For instance, \cite{Chen2015} propose an ad-hoc team formation framework that considers learning other agents' capabilities in the context of unknown tasks. In order to solve a new task, agents would prefer to team up with unknown agents instead of with agents whose known capabilities do not adjust to the task. They observe that learning the capabilities of others in the context of agent and task openness improves team composition and task resolution.  

\subsubsection{Personality: Individual behaviour models} 
Personality is key to understand people's behaviour, cognition and emotion. The use of personality models in agents helps to create more realistic complex scenarios. Indeed, autonomy is related to how individuals behave and what makes them behave differently, even when facing the very same situation. Personality provides a mechanism for behaviour selection that is independent of social background (such as beliefs or morality).
Very recently some MAS contributions have started to consider the notion of personality, i.e. individual behaviour model, to compose heterogeneous teams. For instance, \cite{Hanna2015} study the influence of two agent personality traits: extraversion and agreeableness, both expressed as verbal and non-verbal communication skills. They construct pairs of human users and Intelligent Virtual Agents (IVAs) and analyse how the personality traits influence the development and maintenance of a Shared Mental Model (SMM). The results confirm the importance of providing IVAs with these personality traits to succeed in jointly solving tasks. On a different vein, \cite{Andrejczuk} use personality traits to partition a group of humans into psychologically-balanced and gender-balanced heterogeneous teams with the purpose of increasing the overall performance of the resulting teams.

Marcolino et al. \citep{MarcolinoIJCAI2013,Marcolino2015,Marcolino2016} propose a new approach for action selection. A task is a sequence of actions to be decided at execution time. To choose which action to execute next, every heterogeneous agent within a team votes for its preferred candidate action. Agents vote according to a probability distribution over actions that varies for each agent.  This can be understood as a way of modeling an agent's personality, motivations and beliefs (causing him to behave in a certain way).

  In a series of papers, \cite{FarhangianPPS15,Farhangian2015} use the Myers-Briggs Type Indicator (MBTI) \cite{MyersBriggs} scheme to model different agent personality types. \cite{Farhangian2015} is the only previous work to our knowledge that uses both individuals' skills and personality types (measured by MBTI and Belbin  \citep{Belbin1993} personality tests) to compose teams. These two dimensions are used to simulate human team composition in a business environment.

Another aspect covered by the existing literature is the individual agent knowledge about the other team members' personalities, that is, about their behaviour models. These works go beyond many ``ad-hoc'' team composition systems where information details about the behaviour of individual agents is absent. \cite{Barrett2013} focus on how a new member in a team behaves in order to cooperate well with the other team members whose behaviors are unknown. Each agent is endowed with a learning mechanism for building models of the behaviours of many distinct types of other agents via repeated interactions.  A similar setting is presented by \cite{Agmon2014}, though they consider that there are only two types of agents: a best response agent (choosing his action based on the current state of the world), and an ad-hoc agent (has a better awareness of the team’s possible actions and the resulting joint utility). There is no a-priori model, hence, similarly to \cite{Barrett2013}, an ad-hoc agent needs to decide his behaviour by observing his peers. 

\paragraph{Analysis.} In summary, team composition and formation research has focused so far on cooperative, heterogeneous agents that have a set of properties. These properties can be categorized into two groups: capacity and personality. To our knowledge, besides \cite{Farhangian2015}, there has been no further attempts to combine capabilities and personality for team composition and formation in the area of MAS. Besides that, we observe that the capabilities of agents are always static, but the behaviour model may change with agents' interactions. While the capabilities of humans change over time, the MAS literature typically does not consider dynamic capabilities for software agents. 
Finally, when modeling agents' properties, many existing approaches typically assume extensive a-priori information about teammates. This is a strong limitation for real-life settings. Notice that in many companies there is no central and extensive knowledge about all employees' capabilities.

\subsection{WHAT is the problem? The notion of task}
In its most general sense, a task is a course of action to achieve a goal. The execution of a task is then usually equated to the execution of an action plan. Action plans can be rather complex as they may take into account concurrency of actions, time constraints, action order, or environment uncertainty. However, in the team formation literature it is often the case that simplifying assumptions are made and tasks are assumed to be solved by simple action plans. For instance, an action plan can be seen as a set of actions, or even as a set of competences.  In this latter case the idea behind is that the task can be successfully solved by a team of individuals with expertise in a number of different fields. In this section, we review which concepts of task have been proposed in team formation and team composition. We identify two main approaches:

\begin{itemize}
\item Individual-based, i.e. capacity or personality (see section \ref{skills});
\item Plan-based, e.g. the set of actions or subtasks. 
\end{itemize}

Next we discuss each approach in detail.

\subsubsection{Individual-based approaches} 
Sometimes teams work less effectively than initially expected due to several reasons: a bad balance of their capacities, bad personal relations, or difficult social situations. Hence, in order to make sure a task is performed the most effectively, the large body of literature defines the action plan of the task as a set of requirements for agent individual characteristics. It is assumed that the task can be fulfilled if the task requirements are a subset of the capabilities of team members. We categorise existing work on team composition with the purpose to solve a task into two categories of individual properties: capacity and personality. 

\paragraph{Capacity.} The capabilities of team members are crucial while performing a task. For instance, it is obvious that in order to develop an online Java application, the collective team knowledge has to include Java, Java EE, front-end tools, and database and server knowledge. In the MAS literature (as discussed in  Subsection \ref{capabilities}), the majority of research work expresses capabilities as binary (they are present or they are not)  \citep{Anagnostopoulos12onlineteam,Chen2015,Crawford,Okimoto}. The main shortcoming of the binary approach is the restrictive assumption that if an agent has a capability, his expertise level is sufficient to perform a given task, which implies that the quality of the task performed is not relevant. 

In many cases, the definition of a task is indirectly connected to the agents' capabilities. \cite{JAR2015} propose a model where a task is defined as a tuple that contains the specification of the task (i.e. its subtasks) and the deadline by which the task has to be completed. Each subtask is then matched with one capability. A contract net algorithm is used to compose a team of agents that covers all the required capabilities while maximizing the reputation of the team, thus leading to the best expected performance. 
In \cite{Chalkiadakis2012}, a \emph{project} is defined as a set of tasks, where each task has a complexity level (e.g. moderate or ambitious). Agents' capabilities are graded (e.g. a good carpenter). Tasks are matched with agents' capabilities. The probability of an agent succeeding at performing a task depends on the capability degree of the agent performing the task and the complexity level of the task. These probabilities are learned through repeated interactions between agents, and then used by them to self-organise as teams. 
Finally, in Roles and Teams Hedonic Games (RTHG) \citep{Spradling2013} each agent expresses his preferences over both his own roles within a team and on the set of roles needed in the team. This way, agents themselves jointly select a set of required capabilities to perform a given task.

\paragraph{Personality} 
Similarly, personalities of team members are crucial for performing tasks. According to \cite{Wilde2009}, different types of tasks require different personalities in a team. In detail, people with different personalities approach tasks in a diverse way, resulting in better and faster solutions. Along this line, \cite{Andrejczuk} propose a team composition algorithm that groups agents into different teams so that the personalities in each team are as disparate as possible and gender is balanced.

In \citep{FarhangianPPS15}, the nature (structure) of a task is quantitatively characterized: from extremely structured to extremely open-ended. While structured tasks are straightforward and do not require planning, open-ended tasks require creativity and imagination from team members.
In another article, \cite{Farhangian2015} try to capture the dynamics of tasks by matching the required levels of creativity, urgency, social interaction and complexity of a task to personalities of agents. For instance, teams composed of differing attitude tendencies (associated with different personalities) are believed to outperform  teams composed of like-minded people when tackling tasks requiring a high level of creativity.

Finally, \cite{Hanna2015} show that when performing a task, the personality of team members influences their success. They analyse the influence of an Intelligent Virtual Agent (IVA) communication style (expressing its personality) on human-IVA cooperation. The task is a collaborative game that involves dodging a sequence of obstacles to reach a target.  

\subsubsection{Plan-based approaches}\label{plan}
The notion of task in plan-based approaches is normally understood either as a set of actions or as a sequence of actions.
Well organized teamwork can shorten the time required for completing a particular task by distributing a set of actions across team members.  
Both \cite{Barrett2013} and \cite{Agmon2014} employ an indirect planning method driven by the most informed agents to solve a set of actions.
\cite{Barrett2013} introduce an ad-hoc team agent that learns its teammates' models (i.e. their predictable action selection) and chooses its own actions so that they collectively maximize the likelihood of success. In detail, they use Monte Carlo sampling to simulate the long term effects of collective actions. 
As an extension to the previous work, in \cite{Agmon2014} the actions selected by ad-hoc agents influence the actions that the other team members will choose. Each agent has a set of possible actions that it may choose in order to solve each subtask. The ad-hoc agents need to predict the actions of its teammates (conditioned in this case to its own actions) and behave based on these predictions with the purpose of influencing the collective selection of actions in the team to reach a joint optimal solution. 

Among the approaches considering a task as a sequence of actions, 
in \cite{MarcolinoIJCAI2013} a team of agents jointly playing the computer game Go plan which action to take next by voting on the possible alternatives from a discrete set of possible actions. Authors prove that under certain conditions of opinion diversity, aggregating the decisions of a team of heterogeneous agents is a better planning strategy than the decision of a team built with copies of the most competent agent (called the strongest agent). This shows that diversity improves the planning capacity of a team solving a complex task like Go. 
In \cite{Marcolino2016}, the authors use the same technique to suggest a user a number of optimal solutions for their next action decision.  The application domain of their algorithm is house design. Various design alternatives are proposed to the user in order to select one for further study.

Finally, \cite{Rochlin} deal with self-interested agents in a team that select one agent to accomplish the task of purchasing a jointly desired item with the lowest possible cost. By doing so, the team assigns the execution of the plan to a single member of the team, becoming the buyer. The buyer's strategy decides whether to maintain the search looking for better deals (search for a further action), or stop looking and buy at the lowest price found so far, bearing the incurred buyer's overhead. This strategy balances the expectation of finding a better price (considering the price distribution built during the search) and the team policy to reimburse the cost of the task solution finding to the buyer.   

\paragraph{Analysis.} In conclusion, tasks are solved by the execution of action plans. How complex these action plans are depends on the focus of the reviewed contributions. Individual-based approaches understand action plans as sets of requirements on a team members' capacity and personality. These approaches assume that the joint capabilities of agents in a team must be enough to solve a given task. 
Contrarily, plan-based approaches regard tasks as sets of actions or sequences of actions that are assigned to the individual members of a team. All these works propose algorithms that determine which action will be executed and by whom. However, plan-based approaches have a very simplistic notion of plan. The majority of models do not consider time constraints, action dependencies, action failure, plan robustness, or dynamic changes in a task requirements. Therefore, the vast literature on planning has not yet been integrated into team formation methods. 

\subsection{WHY do we do it? The objective(s)}\label{goal}
The motivation of individual efforts or actions is to attain or accomplish a certain state of affairs: its goal. A necessary condition for a team to exist is that all team members are committed to a joint goal. Therefore, in Computer Science an agent team is typically built of at least two cooperative agents that share a common goal; by teaming up, these goals can be achieved in a more effective way. This is the main motivation of team composition and formation. 
A large body of literature proposes team composition algorithms to attain at least one of the following team objectives: 
\begin{enumerate}
\item minimizing overall cost (e.g. cooperation cost, team cost); 
\item maximizing social utility; or 
\item maximizing the quality of an outcome. 
\end{enumerate}

In this section we describe the literature on team composition per objective.

\subsubsection{Minimizing overall cost} Team cost efficiency has received some attention in the literature. There are various costs associated with team composition and formation problems (e.g. communication costs or agent service costs). For instance, some results balancing cost and quality were obtained by \cite{Kargar}. They propose algorithms for composing a competent team in a social network. When composing a team, those algorithms minimize team members' costs and communication costs within the team. \cite{Kargar} require that agents have the necessary competences to perform a task, but do not require any specific motivation from them.

A similar approach is presented in \cite{Crawford} and \cite{Okimoto}. These works propose a model for robust team composition and go a step further with respect to \cite{Kargar} since they minimize the overall cost among k-robust teams (see Section \ref{teamcompo} for a definition of a k-robust team). That is, this model assumes that up to $k$ agents within a team may eventually fail without affecting the achievement of the task. Thus, it assumes more realistic conditions than \cite{Kargar}. However, likewise \cite{Kargar}, agents' motivations to work together in a team are not considered.

\subsubsection{Maximizing social welfare} A second objective considered in the team composition and formation literature is maximizing social welfare. That is, maximizing the utility function of a team, as a whole, while performing a task. The utility obtained is then allocated to the individual members of the team. For instance, \cite{Chalkiadakis2012} propose a Bayesian Reinforcement Learning framework where agents learn from iterated coalition compositions. Agents can choose between exploration (select coalitions to learn more about new agent types) and exploitation (rely on known agents). Exploitation enables agents to maximize their utility function by performing tasks with reliable agents (discovered during the exploration phase).

Paradoxically, the agent motivation to maximize its individual welfare may reduce the overall team cost and additionally increase the overall quality of the performed task. For instance, in \cite{Rokicki} a human team competition mechanism improves cost efficiency and the quality of a solution in a team-based crowdsourcing scenario. In conventional crowdsourcing reward schemes, the payment of online workers is proportional to the number of accomplished tasks (pay-per-task). Rokicki et al. examine the possibility of getting much higher rewards by introducing strategies (e.g. random or self-organised) for team composition. Their mechanism triggers the competition among human teams as the reward is only given to the top-5 performing teams or individuals. Their evaluation shows substantial performance boosts (30\% in the best scenario) for team-based settings without decreasing the quality of the outcome.

The objective of maximizing social welfare is also considered in many ad-hoc settings, like the one proposed by \cite{Agmon2014}. Agmon et al. consider a framework with two types of agents: best-response and ad-hoc agents forming teams. On the one hand, best-response agents have limited knowledge and assume that the environment and their teammates will behave as observed in the past. On the other hand, ad-hoc agents have a more complete view of a team actions, agents' joint utilities and their action costs. Using such information, ad-hoc agents try to influence joint decisions. In \cite{Agmon2014} the authors consider that ad-hoc agents know with uncertainty their teammates' behaviour. The paper analyses the impact on optimal solutions of ad-hoc agents misidentifying their teammates' types. 


The study of self-interested agents that co-operate in a team has also attracted the interest of researchers in MAS. 
An interesting example of this approach is presented in \cite{FarhangianPPS15}, where  self-interested agents need to maximize the welfare of all team members in order to maximize their own benefit. Hence, they indirectly aim at maximizing the utility of the team. Similarly, in \cite{Chen2015} agents repetitively decide which team to join by balancing both rewards from completing tasks and learning opportunities from more qualified agents. That is, each agent consider whether to sacrifice short-term rewards to acquire new knowledge that benefits himself and the whole community in the long run. 

\subsubsection{Maximizing quality} The last range of models propose a number of methods where agents try to maximize the quality of solutions whilst minimising the time to achieve them, namely to maximize team performance.
Recent organisational psychology studies prove that team members' diversity is a key factor to increase team performance \cite{Wilde2009}.
As mentioned in Section \ref{skills} \cite{MarcolinoIJCAI2013} present a setting where agents in a team vote together to decide on the next joint action to execute that maximises the team's solution quality.  The authors prove that a diverse team can overcome a stronger team (i.e. a team built of copies of the strongest agent) if at least one agent has a higher probability of taking the best action in at least one world state than the probability that the best agent has of taking that action in that state. 
The attempt of capturing heterogeneity is also used in \cite{Andrejczuk}.  There, instead of looking for a single heterogeneous team, Andrejczuk et al. partition a group of agents into psychologically-balanced and gender-balanced heterogeneous teams with the purpose of increasing the overall performance of the resulting teams.

\cite{Hanna2015} also use personality to investigate the influence of Intelligent Virtual Agents (IVA) on team collaboration. Their findings reveal that team performance boosts when the human and the IVA in a team have a shared mental model. Building a shared mental model is directly related to the psychological traits of IVA.

\cite{JAR2015} introduce a decision making mechanism that on top of improving the quality, aims at increasing the quantity of completed tasks. It uses reputation and adaptation mechanisms to allow agents in a competitive environment to autonomously join and preserve coalitions (teams). In terms of team performance, they show that coalitions keep a high percentage of tasks serviced on time despite a high percentage of unreliable workers. Moreover, coalitions and agents demonstrate that they successfully adapt to a varying distribution of incoming tasks.

\cite{Liemhetcharat2012} developed a model to learn and analyze capabilities of agents and synergies among them to solve the team composition problem using previous joint experiences. They define a synergy model as a graph where the distance between agents is an indicator of how well they work together. Their main contribution is that their algorithm learns from only a partial set of agent interactions in order to learn the complete synergy model. In a subsequent article \citep{Liemhetcharat2014}, the authors study the learning agent team formation problem with the goal of maximizing the mean performance of a team after $K$ learning instances. There, learning agent pairs have heterogeneous rates of coordination improvement, and hence the allocation of training instances has a larger impact on the performance of the final team.

The notion of fairness is also considered in the context of team performance. 
An example of this approach is given in \cite{Rochlin}. Rochlin et al. analyze the correlation between efficiency and fairness in teams consisting of self-interested agents. They prove that the more fair the team the more efficient its members are.

\paragraph{Analysis.} In summary, the computer science literature has focused on team co-operation with various objectives that can be categorized as at least one of the following: minimizing overall cost, maximizing  social utility, or maximizing team(s) performance. 
The models minimizing overall cost compose teams based on individual competences, though do not take into account individual motivations to complete the assigned task. This is a rather strong assumption, especially when it comes to mixed teams or human teams, making the existing approaches rather unrealistic. 
The literature focusing on maximizing social welfare considers both agent competences and motivation. The motivation increases by using competence mechanisms (like in crowdsourcing teams), or by giving agents the freedom to select their collaborators (like in learning agent team formation or in ad-hoc teams). 
To maximise team performance, one of the crucial findings in both Organisational Psychology and Computer Science is that team members must be heterogeneous. 
Further variables that have been used by computer scientists in the area of MAS to compose teams are: agent reputation, personality of humans and agents, synergy between team members, and feeling of fairness among team members.

\subsection{HOW do we do it? The organisation} \label{howcs}

In the existing literature, the societal structure of teams is considered crucial for effective teamwork. There are two aspects to be considered, one is which agents will be members of a team and second, how teams will be organized to solve tasks. Thus, the different approaches in the literature can be classified depending on the functionality that they tackle:
\begin{itemize}
\item Team Composition: the process of deciding which agents will be part of a team. It can be an external decision or an autonomous decision by the agents themselves; and
\item Team Formation: the process of learning to decide the roles and internal organisation of a team. This organisation can be imposed or be the result of self organisation. In any case, the resulting organisations can be categorized as hierarchical or egalitarian.
\end{itemize}

Next, we look into these two dimensions in detail.

\subsubsection{Team Composition.}\label{teamcompo} Although team composition in MAS has mainly focused on building teams of software agents, that is agent teams, there is a growing number of works considering either mixed teams \citep{Hanna2015}, where agents and humans cooperate to achieve common goals \citep{Ramchurn2016}, or human environments, where people are supported by software \citep{Jennings2014}. In MAS, we distinguish between two groups of methods (or processes) to compose team(s), namely:
\begin{enumerate}
\item Exogenous Team Composition: there is an algorithm external to the agents that determines the composition of teams.
\item Endogenous Team Composition: agents themselves decide in a distributed manner the composition of a team.
\end{enumerate}

\paragraph{Exogenous Team Composition.} The team composition process uses the task requirements (i.e. constraints on teams that can be formed, such as team size \citep{Rahwan2011}; competences and personality as discussed in section \ref{skills}) in order to build teams that are capable of solving the task with particular properties. For instance, \cite{Crawford} and \cite{Okimoto} consider a degree of fault-tolerance to build $k$-robust teams. A team is $k$-robust if removing any $k$ members from the team, does not affect the completion of the task. 
As mentioned before, \cite{Liemhetcharat2012} propose a learning algorithm that constructs a synergy graph from observations of the performance of pairs and triples of agent. A synergy value represents how well a pair of agents work together. The authors use this learned synergy graph as well as agent capabilities to solve the team composition problem. Their method selects teams that are capable and that maximize their internal synergy. 

Similarly, \cite{Rangapuram2015} consider the competences of agents and their compatibility in order to identify a team that is both competent and compatible. Agent compatibility, expressed as a social network, can be understood as a set of preferences on team composition, such as: the inclusion of a certain team leader, or restrictions on team size, problem solving cost or agent locality (in a social or geographical sense).

In many systems, capabilities of agents are not widely known. \cite{Chen2015} study an ad-hoc setting where agents need to co-operate with to recognize their capabilities. Agents bid for subtasks (parts of tasks) that they want to perform, though the final decision belongs to the exogenous algorithm that assigns each subtask to the best qualified agent bidding on the task. 

Some approaches deal with the composition of multiple teams. For instance, \cite{Anagnostopoulos12onlineteam} use competences and communication cost in a context where tasks sequentially arrive and teams have to be composed to perform them. Each task requires a specific set of competences and the team composition algorithm is such that the workload per agent is fair across teams. Furthermore, \cite{Andrejczuk} compose multiple teams according to a balance of agents' personalities and genders. Their goal is to partition a set of agents into multiple teams such that each team is internally balanced and the problem solving capabilities of the teams in the partition are similar. Besides the use of personality traits, \cite{Farhangian2015} use competences and a task specification with the purpose of composing a single team. 

Aside from competences and personality, team composition can also take into account agents' preferences on teams. Indeed, hedonic coalition formation employs each agent's \emph{hedonic preferences} on its coalitions to yield a coalition structure, namely multiple teams. The defining feature of a hedonic preference is that every agent only cares about which agents are in its own team (coalition). \cite{Spradling2013} introduce a new model of hedonic coalition formation game, the so called Roles and Teams Hedonic Games (RTHG). In this model, agents view coalitions as a number of available roles and have two levels of preferences: on the set of roles that are available in a coalition, and on their own role within each coalition.



Finally, there is recent, relevant work on mixed teams by \cite{Hanna2015}, which composes a team as a pair consisting of a human and an Intelligent Virtual Agent (IVA). The pair play a collaborative game that involves passing a sequence of obstacles to reach a target.

\paragraph{Endogenous Team Composition.}\label{endo} The second group of methods for organizing teams has an endogenous nature. They incorporate algorithms enabling agents to decide on team composition by themselves. In detail, agents are equipped with negotiation and decision-making mechanisms that they employ to agree among themselves on a team structure. Therefore, team composition occurs without explicit external command. 



\cite{FarhangianPPS15} propose a model in which there are two types of agents: requesters in charge of tasks that seek for contributors to compose teams, and contributors that vote for the tasks they want to perform.  Each requester runs an auction-based (first-price sealed-bid) algorithm with the purpose of composing teams with the highest chance to increase social wealth. Contributors issue bids pursuing to join the most useful requesters, namely the ones that are most likely to reward them. \cite{JAR2015} follow the similar approach but also employ reputation and adaptation mechanisms to allow agents in a competitive environment to autonomously join and preserve teams (as coalitions). Agents bid for tasks and each team is constructed and led by a mediator agent. 

Similarly, in \cite{Chalkiadakis2012} each agent builds its beliefs about its peers based on prior outcomes of interactions between them, and decides on coalitional actions (which coalition to join and what task to perform). Then, agents negotiate between them to form teams taking into account their own beliefs on the probability of success when being in a team. 

There exist also mixed approches, where researchers explore both, exogenous and endogenous methods to compose teams. For instance, \cite{Rokicki} propose strategies for groupsourcing (team-based crowdsourcing), ranging from team formation processes where individuals are randomly assigned to teams, to strategies requiring self-organisation where individuals participate in team building. Their results show that balanced teams (that is teams with the balanced number or agents in each team) combined with individual rewards for most effective team members outperforms the other strategies.

\paragraph{Analysis.} The majority of researchers focuses on exogenous methods to compose teams. However, there are many actual-world application domains (e.g. co-working, or crowdsourcing) where endogenous team composition and formation are more appropriate for deployment.

Most of the literature on exogenous team composition assumes that there exists a centralized, detailed knowledge about all agents. This knowledge is required in order to compose teams based on agents' capabilities, personality, or even preferences. Endogenous methods are best for dynamic environments, where team composition and formation processes are continuously performed. Furthermore, it is a good setup for agents that learn other agents' capabilities through repeated interactions.


\subsubsection{Team Formation} We identify two main team organisation structures to build effective teams:
\begin{enumerate}
\item Hierarchical; and
\item Egalitarian.
\end{enumerate}

We describe each team organisation structure in the following sub-sections.

\paragraph{Hierarchical.} A hierarchical structure considers a team leader who is responsible for and makes the decisions affecting the team. This structure is the traditional setting when it comes to business units.

As mentioned in subsection \ref{endo}, \cite{FarhangianPPS15} consider two types of people within teams: requesters and contributors. Requesters adopt a leading function, they start a project and recruit the required people. Contributors perform the tasks assigned by requesters. The overall team behaviour is determined by the personality of agents in teams.

In \cite{JAR2015}, each coalition is led by a mediator. This agent is responsible for leading a coalition by selecting suitable agents to be part of a coalition (called worker agents) and by evaluating the performance of workers while the coalition operates.

\cite{Agmon2014} consider ad-hoc settings with two types of agents: best-response agents and ad-hoc agents. In such settings a task consists of a set of actions, and each team becomes responsible for performing a task. Each best-response agent selects its next action based on its own local world view. Each ad-hoc agent acts to bring out the best in its teammates by ``leading'' them to the optimal joint action. This is an arresting example of a hierarchical structure, where agents are not aware of each other's roles, and hence of a team's structure. Nonetheless, an ad-hoc agent has more knowledge than a best-response agent, and thus it exploits such information to lead its team. This may happen in a business setting, where both senior and junior staff form a team. Even though there is no clear division of roles, the senior
employee uses his experience to make decisions that are best for the team in a long--term period (and may not look best from a short--time perspective).

\paragraph{Egalitarian.} An egalitarian structure assumes that all workers in a team are equally informed and 
have the same rights. The leadership within a team is shared and existing team roles result from the team's task requirements. An example of this structure in real-life scenario might be a team of doctors that need to join their specialized knowledge to perform a complicated surgery on a patient.

A large part of the MAS literature focuses on the egalitarian setting, trying to benefit from leaderless teams that cooperate to complete tasks. We find this team structure in Groupsourcing \citep{Rokicki}, Robust Teams \citep{Okimoto,Crawford}, Ad-hoc teams \citep{Chalkiadakis2012,Chen2015,Barrett2013}, Mixed Teams \citep{Hanna2015}, or Learning Teams \citep{Liemhetcharat2012,Liemhetcharat2014}.



A particular case of egalitarian structure involves  members that decide collectively, usually by voting, on the appropriate course of action while performing an assigned task. The real life example for this organisation structure might be a start-up with few people that make all decisions by discussion. 
\cite{MarcolinoIJCAI2013,Marcolino2015} and \cite{Marcolino2016} study egalitarian structures whose agents vote to decide at every step of a task in order to choose the best course of action. They prove that teams consisting of heterogeneous agents that vote their actions are more efficient than homogeneous teams built out of the copies of the strongest agent in a team. This is because the spectrum of possible actions is wider for heterogeneous teams.

Finally, some team composition models can produce both types of team structures. For instance, in Roles and Teams Hedonic Games model \citep{Spradling2013}, the resulting structure of the teams can be either hierarchical or egalitarian depending on the relationships between roles. Typically teams in \citep{Rangapuram2015} are egalitarian, though the presented model includes many natural requirements that can lead to a hierarchical structure (such as inclusion of a designated team leader and/or a group of given experts).

\paragraph{Analysis.} The team organisation structures in the MAS literature can be grouped into hierarchical and egalitarian. The majority of MAS research focuses on egalitarian structures because of simplicity reasons. In particular, there is no need for defining a role structure together with its relationship and agent-role assignments. 
Although structuring teams and organisations largely helps reduce complexity of interactions, by separating responsibilities, most research in team formation does not consider a clear role division. 
Moreover, notice that in most business settings teams work following a hierarchical structure.


\subsection{WHEN do we do it? The dynamics}
The literature on team composition and formation mostly considers that tasks are static in the sense that their requirements do not change during their execution. However, the dynamics of task arrival is considered by many. That is, there could be multiple tasks to be solved concurrently and new tasks may arrive in an asynchronous, localized manner. The different works consider different issues in this dynamic process. For instance, the number of tasks to be serviced, task and team members localization, team size per task or time limitations. Normally, if there is only one task is to be completed, the focus will be on composing the best team for the task. On a repeated task arrival setting, the use of a history of team work experiences is key to compose new teams.  
Hence, the literature can be classified depending on two main aspects: 
\begin{enumerate}
\item The succession of  tasks,
\item The simultaneity of tasks.
\end{enumerate}

The simplest case is a one-shot task. There is neither succession nor simultaneity, and hence the problem of team composition is normally reduced to finding the best team for the only task. When tasks come in sequence without simultaneity, then the problem can be reduced to finding the best team for each task while using the learned experiences in the composition of each new team. If  tasks come in succession and can be simultaneous, the need to deal with multiple teams acting at the same time becomes a key issue. The succession of possibly simultaneous tasks is the most complex framework in which memory becomes again a key element.

We discuss each aspect in detail.

\subsubsection{Non Successive and non simultaneous tasks}\label{nonnon}
In this case we face a one-shot task resolution. This is the simplest case for the team composition and formation problems. There is no long-term strategy used to compose and form teams. Thus, agents do not learn from past experiences and we cannot talk about the notion of community in this setup. 

\paragraph{Team Composition.} As mentioned above, in the team composition problem, we are looking for only one team, the best possible one to perform the task. The majority of models that consider non successive and non simultaneous tasks are simplistic. They assume that once the team is composed it has the needed properties and will perform the task well. For instance, \cite{Kargar} use agents' capabilities and team coordination cost to compose the most effective team. Similarly, \cite{Crawford} and \cite{Okimoto} use agents' capabilities to compose k-robust teams (see Section \ref{teamcompo} for a definition of a k-robust team). In \cite{Rangapuram2015}, besides agents' capabilities, the team composition model also introduces various types of constraints (the inclusion of a specific group of agents in a team, team size, budget limitations, and maximum geographical distance between agents and between agents and tasks). This last model is more realistic, though it disregards past experiences.

\paragraph{Teamwork.} In the teamwork phase, agents solve the task once and for all.  Hence, one-shot tasks may cause self-interested behaviours, such as in \cite{Rochlin}. There, as mentioned in Section \ref{plan}, one agent (called buyer) from the team is delegated to accomplish the task of purchasing a jointly desired item with the lowest possible cost. This agent operates on a one-time setting, that is, there is a single agent deciding on behalf of the team, and hence, there is no need for that agent to behave in an altruistic manner. Authors study the notion of fairness and its influence on effectiveness. They show that the selected buyer is less motivated to do the task if the cost of the goods is to be divided equally among the team members. In this case, the purchasing costs are fully assumed by the purchasing agent. Therefore, they study different methods to reimburse the purchasing costs incurred by the buyer to improve its effectiveness. 

\cite{Hanna2015} study the co-operation between a human and an IVA (Intelligent Virtual Agent) in a one-shot task setting. Given that past experiences cannot be used, they experimentally show, by comparing many one-shot task instances, that the more informative the communication between the two agents, the better the performance of the team. The communication behavior of an IVA is directly related to its psychological traits.

On a different vein, many models assume that given a one-shot task, agents will behave according to their knowledge and capabilities in order to benefit the whole team. In \cite{Barrett2013} and in \cite{Agmon2014}, team agents are pre-designed to co-operate when solving a collective task. Then, one of the agents is replaced by an ad-hoc agent that shares the team’s goals, though does not know its teammates’ behaviours. The ad-hoc agent cannot control its teammates, and yet it tries to improve the team’s performance by learning to predict other agents’ actions and thus selecting its own actions to achieve an overall optimal team behaviour. \cite{MarcolinoIJCAI2013} and \cite{Marcolino2015} perform a one-shot task study, where team agents vote for a team action leading to the task resolution. The action voted for is sampled from a fixed probability distribution over those actions appropriate in a particular world state (no learning involved). The higher the probability of an action the more preferred it is by the agent. A plurality voting mechanism is used to select the team action. Authors show that a diverse team (with different probability distributions) can outperform a uniform team (made out of copies of the best agent) and that breaking ties in favour of the best agent's opinion  in a diverse team is the optimal voting rule \footnote{Notice though that the authors make the strong assumption that there is a known rank of the best actions to take at any time.}.

\subsubsection{Non Successive and simultaneous tasks}
In non successive and simultaneous tasks, the composition and formation problem becomes more complex as it now considers a set of one-shot tasks. There is still no use of the past experiences as the tasks are non successive.

\paragraph{Team Composition.} Researchers in the area of MAS propose algorithms to compose the \emph{best set} of teams, one per simultaneous task, instead of looking for the \emph{best} team for a task. For instance, \cite{Andrejczuk} partition a set of agents into gender- and psychologically-balanced problem-solving teams that have to solve different instances of the same task. The authors use a greedy technique to balance the psychological traits of the members of teams so that each team gets the full range of problem-solving capabilities.

In Roles and Teams Hedonic Games (RTHG) \citep{Spradling2013} authors propose a heuristic optimization method to partition a set of agents, again to solve different instances of the same task. The method treats as votes agents' role preferences on team role structures. Firstly, the role structures of the teams will be those receiving the highest social welfare (as the summation of the agent individual utilities to play any of the roles in the structure). 
Secondly, the algorithm selects the agent with the highest utility for a remaining role in the most voted team role structure, recomputes the role structure preferences without that agent's preferences, and keeps staffing teams until the partition is complete.
For instance, an agent may prefer to be a programmer in a two-agent team including a designer, but would not like to play any role in a team without a designer. 
Hence, an agent’s role preference is not taken in isolation, but in the context of the teams' composition. Authors define Nash stable and individually stable solutions for RTHG in terms of possible local moves that agents could make within a given coalition partition and prove that every instance of RTHG has an individually stable partition that can be obtained with the use of local search movements (change of role within a coalition or coalition swaps).
In our literature search, we could not find approaches dealing with different simultaneous non successive tasks.

\paragraph{Teamwork.} Similarly to team composition, \cite{Rokicki} deal with the Teamwork problem over different and simultaneous instances of the same task. Agents may change their strategy during team formation in order to reach a better solution. They classify human behaviour during team self-organisation in crowdsourcing tasks in two types. First, a number of users choose to join one of the leading teams, instead of selecting a weaker one and compete for a lower award. Second, small teams merge to form stronger teams and thus have a higher chance of achieving an award.

\subsubsection{Successive and non simultaneous tasks} 
When tasks are successive and non simultaneous, the algorithms for team composition and formation deal with a task that has to be assigned to a team, and in many cases solved, before new tasks arrive.
A successive setting can discover phenomena which we believe are important, but which are not captured when the attention is limited to static, non successive tasks. If in the system of the same set of agents, teams are created and dismantled depending on the task, the agents may behave very differently than in a non successive settings. For instance, a person will behave in a different manner if she repeatedly borrows a car from her friends, than when she simply rents a car. The successive setting has its advantages: it lets agents learn from the past experiences and build their beliefs based on this knowledge. 

\paragraph{Team Composition.} 
In \cite{Anagnostopoulos12onlineteam}, the first task arrives at the first time step and is assigned to a newly composed team of experts before the arrival of the second task. This procedure repeats until all tasks are assigned. Authors propose an algorithm to compose a set of teams to handle a set of these incoming tasks. The goal is to form a new competent team upon arrival of each task, so that the workload in the whole system is balanced. There is no learning involved in this process. Contrarily, in \cite{Liemhetcharat2012} a learning algorithm is proposed that constructs a synergy graph from observations of the performance of pairs and triples of agent in solving previous tasks. The synergy tells how well a pair of agents work together and they use this learned synergy graph as well as agents' capabilities to solve the team composition problem for the next task. Their method selects teams that are capable and maximize their internal synergy. 

\paragraph{Teamwork.}	
To the best of our knowledge, there are no contributions on teamwork that consider successive and non simultaneous tasks. 


\subsubsection{Successive and simultaneous tasks}
When tasks are successive and simultaneous, the algorithms for team composition and formation deal with a set of tasks arriving, possibly overlapping in time that have to be assigned to newly composed teams.

\paragraph{Team Composition.}
In \cite{FarhangianPPS15}, tasks arrive in any order, possibly overlapping in time. A team is composed for each incoming task and after execution agents assign performance values to each one of the other team members. These values are public and used by the community to compose teams for future tasks. 
\cite{Chalkiadakis2012} present several learning algorithms to approximate the optimal Bayesian solution to the repeated team composition.
Similarly, \cite{JAR2015} compute, after teamwork, both individual agent and coalition (team) reputation values to be used in the composition of future teams. 

Finally, in \cite{Chen2015}, for each new task arriving agents decide which team to join balancing exploitation (rewards from completing tasks learned from previous task solving) and exploration (learning opportunities from more qualified agents leading to future rewards). 

\paragraph{Teamwork.}	
To our knowledge, there are no contributions considering successive and simultaneous teamwork.

\paragraph{Analysis.} One time settings (i.e. non successive tasks) are usually simplified models that do not take into consideration the history of agent interactions. One-shot tasks may cause self-interested behaviours, where agents look for at least a fair split of costs associated with teamwork. However, the majority of the literature on team composition and teamwork considering this setting assume that the agents will always behave accordingly to their capabilities and knowledge.
The successive tasks provide us with more realistic and complex scenarios. The tasks arrive either in order, one after another, or overlapping in time.  The majority of the literature uses this setting to let agents build their beliefs based on the past experiences and compose new teams according to these beliefs. 
Regarding teamwork, there are no contributions that explore successive settings. In other words, the state of the art does not acknowledge the memory of agents as important while executing tasks.


\subsection{WHERE do we do it? The context}

The context is understood as the circumstances that form the setting for the team composition and formation processes. We observe that the concept of context in the reviewed computer science literature has not played a major role so far. Contrarily, according to the organisational psychology literature \citep{guzzo1996}, it is one of the most important variables while composing and forming teams (see Section \ref{whereOP}). There are different categorizations of context. One of them is proposed by \cite{kozlowski}, which classifies contexts as follows:
\begin{itemize}
\item Organisational Context: technology used, organisation structure, leadership, culture, and climate.
\item Team Context: normative expectations, shared perceptions, and compatible knowledge  (generated by and emerge from individual interactions).
\item Individual Context: attributes, interactions, and responses.
\end{itemize}

In the MAS literature there are very few works that consider the social context while composing teams.
In \cite{Rangapuram2015}, while composing teams, the context is exemplified as a social network that  encodes the previous collaborations among experts. The idea behind it is that the teams that have worked together previously are expected to have less communication overhead and work more effectively as a team. Similarly, \cite{JAR2015} propose to express social context by the reputation measure. There, upon task completion, the contractor rates the quality of the service provided by a team and, also teams rate their own workers. Finally, this rating information is maintained and aggregated by a reputation module. \cite{Liemhetcharat2012} propose to model a social context by using the learned synergy graph (that measures how well agents work with one another) and hence, solve the team composition problem. 
\cite{Anagnostopoulos12onlineteam} include the coordination costs by means of a social network over the set of agents and assume a metric distance function on the edges of the network. On top of modeling preferences based on social context (such as past interactions, compatibility in collaborating, distance in a company’s hierarchy), the function may include any other kind of context, (for instance geographical proximity between agents or between task and agents within a team). 

\paragraph{Analysis} To the best of our knowledge, there are only few works in MAS literature that  recognize the context as an important variable. Besides \cite{Anagnostopoulos12onlineteam}, which  considers both social and geographical contexts, the methods in the literature only consider the social context (if analyzed at all).

\section{Team composition and formation from an organisational psychology perspective.}\label{OP}

In this section we discuss all above aspects in detail answering the questions asked in the introduction of this paper.

\subsection{WHO is concerned?} \label{skills2}
In this section we are going to survey the literature on Organisational Psychology that deals with the characteristics of humans composing teams. 

We will use the structure as in section \ref{skills}.

\subsubsection{Capacity.}
In OP, the most important capacity of team members that is related to team performance is their cognitive ability. Hence, the main goal is to study how cognitive abilities influence team performance. 
Cognitive ability refers to the `capacity to understand complex ideas, learn from experience, reason, solve problems, and adapt' \citep[p.507]{devine2001smarter}. Hence, cognitive ability in OP is a much wider concept than capacity in multiagent systems as on top of skills widely used in MAS systems, it contains many other properties such as experience, competences, age or even gender. 

Moreover, in contrast to computer science, where capabilities are static, psychologists deal with the dynamism of human capacity. Humans learn new capabilities and increase their level every day for whole live (see more in \cite[p.399-403]{laal2012lifelong} for the concept of the lifelong learning).  There are diverse tests and methods to examine humans capacity, such as: intelligence or cognitive competences tests, assessment centers or social and behavioural competence tests. 

Regarding team composition, on the one hand \cite{Bell2007} and \cite{devine2001smarter} found that mean team values of cognitive ability are correlated with team performance. Moreover, she also found that the lowest and the highest team members' cognitive abilities are correlated with team performance in lab and field settings.  In addition, \cite{devine2001smarter} found that the variance of team members' cognitive ability did not help predict team performance. These authors also found that the mean value is twice more informative in predicting than the lowest and the highest member’s scores. 
On the other hand, \cite{devine2001smarter} found that cognitive ability influences team performance differently depending on contextual variables (such as working normative procedures or human resources policies). These findings suggest that, when composing a team, organisations and managers should not only take into account the members' cognitive ability, but also the context in which the team will operate. This will be further discussed in Section \ref{whereOP}.  Other researched individual characteristics like the effect of age and gender have produced some mixed results when analyzing their relation with performance \citep{chmiel2008introduction}. Diversity is needed for innovation but can cause as well conflict and imbalance \citep{unsworth2000teams}.

Finally, similarly to computer science literature, the concept of team properties is understood as a sum of humans' individual properties. 

\subsubsection{Personality}
In addition to the before-mentioned individual properties, the literature has examined the role of personality.  
The most prominent approaches have been  the ``Big Five'' personality traits theory \citep{Mount}, Schutz's theory of fundamental interpersonal relations orientation (FIRO) \citep{Schutz} and the Myers Briggs Type Indicator method \citep{White}. 
They have been used to find the personality traits and types associated with team performance. Regarding the ``Big Five'' theory, meta-analytic research has found that certain levels of conscientiousness, openness to experience and agreeableness are good performance predictors \citep{Mount}. 

Another approach is that of the theory of fundamental interpersonal relations orientations (FIRO) \citep{Schutz}. The idea is that humans have several needs (i.e. need for inclusion, control and affection) and that groups with team members that have compatible needs will perform better than those with incompatible ones. Nevertheless,  research has found mixed support for this theory \citep{west2012}.

Some companies have also tried to base their team formation on cognitive styles of the members, by using the Myers-Briggs Type Indicator (MBTI) assessment instrument |\citep{MyersBriggs}, which is a questionnaire that measures cognitive styles along four dimensions: Extraversion --- Introversion, Sensing --- Intuition, Thinking --- Feeling, and Judging --- Perceiving. Nevertheless, there is not enough rigorous research evidence showing its relationship with team performance \citep{west2012}.

There are also novel approaches created with the purpose of team composition and formation. For instance, the Post-Jungian Personality Theory, which is a modified version of (MBTI) \citep{Wilde2013}. It operates on the same dimensions as MBTI. The main novelty of this approach is its use of the numerical data generated by the instrument \citep{Wilde2011}. The results of this method seem promising as within a decade this novel approach tripled the fraction of Stanford teams awarded national prizes by the Lincoln Foundation \citep{Wilde2009}. However, the method is not properly validated and tested, which makes it disregarded by psychologists.

\subsubsection{Analysis.} 
Several correlations have been found between cognitive ability and team performance. The personality is also present while composing teams, although the correlation between personality and team performance is not clearly explained. The most widely used test to measure personality is the ``Big Five''.
Organisational Psychology studies show that besides cognitive ability and personality,  experience and gender are further properties to consider for team composition \citep{West}. 
Indeed, research findings on this topic suggest that diversity in those characteristics can have an effect on team performance and innovation \citep{West}. 
Additionally, some further research has also paid attention to values and has found collectivism and teamwork preferences \footnote{Teamwork preferences refer to team members preferences on other team members to work with.} to be additional good team performance predictors \citep{Bell2007}. 

\subsection{WHAT is the problem?} \label{task2}
When it comes to team composition, the organisational psychology literature has focused on defining task classifications. These classifications have been employed to study the relation between task types and team performance. 
Hence, in this section we will review the most known task classifications and its influence on team performance.

Two of the most widely discussed task classifications are those of \cite{mcgrath1984groups}, \cite{hackman1990groups, hackman1971} and \cite{hackman1975}. While the classification of \cite{mcgrath1984groups} is based on the cognitive requirements of tasks, the classification in \cite{hackman1990groups, hackman1971, hackman1975} is based on the motivation characteristics of tasks (i.e. autonomy, task variety, task significance, task identity and task feedback). The research on team composition show that the classification based on the motivation characteristics predicts more accurately the team performance \citep{podsakoff1997moderating}.

\cite{hackman1990groups} defines a task classification based on motivational requirements composed by seven work task types: 
\begin{enumerate}
\item top management;
\item task force; 
\item professional support task;
\item performing task;
\item human service task;
\item customer service task;
\item production task.
\end{enumerate}
The classification of \cite{mcgrath1984groups} based on cognitive requirement proposes three dimensions that characterize each task type: 
\begin{enumerate}
\item Choose-Execute;
\item Conceptual-Behavioral;
\item Conflict-Cooperation.
\end{enumerate}
Technically speaking each task type becomes a 3-tuple with qualitative values for each dimension. For instance, a routine task would be very executive, medium behavioral and low conflicting. 

After analyzing seventeen classifications in the literature \cite{wildman2012task} came out with a different classification as follows:
\begin{enumerate}
\item Managing others;
\item Advising others; 
\item Human service;
\item Negotiation;
\item Psychomotor action;
\item Defined problem solving;
\item Ill-defined problem solving.
\end{enumerate}
As an alternative perspective, \cite{NavarrodeQuijanoBergerMeneses} propose a task classification based on the task context (namely task complexity, interdependencies between subtasks in a task, and uncertainty about the dynamics of the environment where the task is executed and the lack of information). Their results show that in order to achieve acceptable performance, the greater the complexity, interdependence and uncertainty, the stronger the requirements on the maturity of teams (e.g. joint experience, cohesion) and on the diversity of team members' capabilities. 
For instance, to carry out highly interdependent tasks, all team members should possess coordination skills (maturity) and some of them the capacity to take decisions (diversity). 
Taking into account other task context characteristic (i.e. uncertainty and interdependence) their study results show, the greater the uncertainty and interdependence of task types, the more diverse the competences for team members to cope with complexity.
From the other hand, if the team is overqualified for the task to perform, the motivation of team members decreases and the quality of the outcome is lower or the task is not completed at all.


\subsubsection{Analysis.}
The OP literature provides many different classifications of task types, where the most important are the classifications based on the motivation of individuals, the cognitive abilities and the task context.
Provided the amount of classifications and the apparent lack of consensus among them, we believe that choosing among the several classifications previously presented in order to apply them to the study of team composition is a hard decision. Nevertheless, such decision must be made in order to move forward with the understanding of how a task type can influence team composition. In an attempt to advice researchers, notice that the research show that the classification based on the motivation characteristics predicts more accurately team performance.

From OP perspective team performance cannot be assessed by simply measuring how long it takes for the group to finish a certain task or by counting the number of right answers to predefined and clear questions, which is a common approach in computer science. OP rather analyzes joint team objectives and the team composition and formation setting (such as not realistic deadlines, a number of individuals in a team, the level of stress in a team or the quality of the outcome).

The current research on organisational psychology focus has moved from task analysis so not many results are present. Although task types are defined, different task instances constantly appear because of technological development. That makes it very difficult to keep the pace. That is why the focus on OP moved to competences (understood as cognitive ability, see Section {skills2}). This is why not much work has appeared after defining task taxonomy. At the same time task complexity increased and hence, teams are getting more and more important. Moreover, a clear mapping between cognitive ability of individuals and task types is needed. As a major benefit such mapping would ease team composition.

\subsection{WHY do we do it?} 

In OP the main objective for team composition and formation is to maximize team performance. When measuring it, the research on OP suggests that we should go beyond mere economic criteria, the quality of decision-making processes or other traditional performance indicators \citep{Komaki,Hackman2002}. 

An important difference with respect to the computer science literature is that team performance is considered from two perspectives: objective and subjective. On the one hand, objective team performance refers to the features of the outcome of a team (e.g. quality, delivery time, cost, sustainability). On the other hand, subjective team performance refers to the quality of human resources in a team (e.g. motivation, satisfaction, commitment, illness rate, stress) \citep{quijano2008human}. Therefore, while the first one refers to the delivered output of a team (what customers obtain), the latest one focuses on the inner development of team members. Objective and subjective team performance are significantly correlated (e.g. \cite{quijano2008human}). Therefore, and not surprisingly, the organisational psychology literature considers both types of performances when tackling team composition and team formation (e.g. \cite{MenesesNavarro}). 
The subjective and objective performance of a team are determined by the several aspects of the context (discussed in Section \ref{whereOP}), together with individual characteristics, the task and the team processes. Following \cite{NavarrodeQuijanoBergerMeneses} the subjective and objective performance of a team are determined by the adjustment between the maturity level of the team (e.g. in terms of group development, potential, etc.) and the groups tasks characteristics. 

\paragraph{Analysis.}
An important difference with respect to the computer science literature is that team performance is considered from two perspectives: objective and subjective. Objective and subjective team performance are significantly and directly correlated. Therefore, and not surprisingly, the organisational psychology literature considers both types of performances when tackling team composition and team formation.
The computer science literature can benefit from the concept of subjective team performance that currently disregarded. Therefore, current team composition models, which mainly focus on the objective team performance, need to be extended.

\subsection{HOW do we do it? The organisation}\label{howop}
Similarly to Section \ref{howcs} on computer science, we divide the organisation into two aspects: team composition and team formation.

\subsubsection{Team Composition.}

The organisational psychology research on team composition has been very influenced by task classification. For several authors, there is a relationship between task type and team type (structure). 
For example, according to \cite{hackman1990groups}, there are seven team types based on the task type to perform:
\begin{enumerate}
\item top management;
\item task force;
\item professional support;
\item performing groups; 
\item human service;
\item customer service; 
\item production teams.
\end{enumerate}

\cite{devine2002review} and \cite{delgado2008teams} highlighted that team performance depends on a good matching between team type and task type.

On the other hand, there are multiple team type classifications in the literature based on other criteria \citep{devine2002review,marks2001temporally,gibson1999our}: motivation-based, cognitive-based or context-based (see section \ref{task2}), though none of them has been widely used or accepted. Also, there is agreement that team diversity must be exploited while composing teams. Diversity refers to the degree or level to which the members of a group differ or contrast in one or more properties. Diversity has been shown to have an impact on team performance \citep{Mathieuetal}. 
In their review, \cite{Mathieuetal} point out the vastness of the literature featuring team diversity and draw attention to four main diversity dimensions: demographic, personality, functional background, and attitudes and values. 

\cite{Horwitz_2007} conducted a meta-analysis to understand the relationships between team diversity and team performance. For this, they differentiated between two classes of diversity: bio-demographic and task-related. The former refers to diversity in individual properties that are observable and not learned (e.g. personality, gender, age, ethnicity), whereas the latter regards diversity in acquired capabilities, such as education or expertise. Using meta-analytic techniques, they found task-related diversity to be positively correlated to both qualitative and quantitative measures of team performance. However, they did not find a clear relationship between bio-demographic diversity and team performance. 
Although pointing out the small number of studies supporting these latest findings, their preliminary results seem to give more importance to the diversity of acquired team member properties, such as the type of education or knowledge expertise.

Finally, another factor influencing team performance is team size. Among researchers the size of a team is one of the most frequently studied parameter when analyzing team performance. There is a disparity in the literature due to the fact that appropriate team size is dependent on the task and the social context in which the team operates.
Some studies have found team size to be unrelated to performance \cite{Martz} or that increasing team size actually improves performance without limit \cite{Campion}. However, other studies show that there is an inverse relationship between the size of the team and its performance \citep{Oyster,Bartol}. 

Among others, \cite{Oyster} and \cite{Bartol} show that team size is important when analyzing team performance. Yet, they have offered different recommendations concerning the best size for various types of tasks to achieve acceptable performance. \cite{Oyster} states that the right number of people in a team depends on the kind of tasks team members need to perform. They believe that for teams ranging from four to six, all the team members' competences can be fully used, but for larger teams some members' competences are under-used and this provokes that teams split up. According to the studies of \cite{Bartol}, the optimal number of members for problem-solving tasks is five. He states that there is a limit to the team size, which, if exceeded, causes a drop in the performance of the team. \cite{Bartol} says that in the case of a team containing more than six people there is a tendency to split the team into two, which brings about negative effects. The cause is twofold: high coordination cost and loss of motivation by team members \cite{Oyster}.


\subsubsection{Team Formation.} 
Once a team has been composed, there are different processes that the team carries out to execute the task and achieve the collective goal.
Several classifications of team processes have been proposed in the literature, from which, the most recent and overarching one is the one proposed by \cite{marks2001temporally} and \cite{salas2005there,Goodwin}. Typically the research investigated the ways of implementing team processes and of measuring how well teams perform. 
To begin with,  \cite{marks2001temporally} distinguish between three broad types of processes: action-orientated, transition-orientated and interpersonal. The first ones refer to actions that team members undertake to accomplish goals, namely team monitoring, systems monitoring, monitoring progress towards goals and coordinating activities. Regarding transition-orientated  processes, these are actions related to planning and/or evaluating in order to guide in attaining team goals, that is goal specification, mission analysis, formulation and planning, and strategy formulation. Finally, interpersonal processes are those intended to manage interpersonal relationships. They comprise motivating/confidence building, conflict management and affect management \citep{marks2001temporally}. On the other hand, \cite{salas2005there} built upon previous research and narrowed down the main processes into ``Big Five'' team processes: team orientation, backup behaviour, team leadership, adaptability and mutual performance monitoring.

Another important aspect is that team climate influences the effectiveness of processes. A team climate is defined by the degree to which a team of persons possesses certain core properties that are needed for the team to work effectively. These properties include the interrelationship among team members, the identification of each person with the team and its social values, the coordination of team resources, behaviours and technologies, as well as the desire of each team member to achieve the objectives of the team \citep{MenesesNavarro}. A good climate assures the sharing of resources, mutual rewards and information exchange. It promotes a high level of openness, safety, and a mix of upward, downward and horizontal communication processes that help to increase team performance \citep{KozlowskiIlgen,MathieuMaynardTaylorGilsonRudy,RicoAlcoverTabernero,Knapp}.

A team climate that is conductive to learning requires shared perceptions of work settings \citep{JamesChoiKoMcNeilMintonWright,Brodbeck2003,RamirezHellerBergerBrodbeck}.
According to \cite{Brodbeck2003} and \cite{RamirezHellerBergerBrodbeck}, a team climate conductive to learning is characterized as one in which: 
\begin{enumerate}
\item There is empathy, support, as well as a common understanding among its members, conveying an atmosphere of mutual trust, 
\item There is a regular contact as well as  informal and formal communication processes among its members, 
\item There exists a common agreement with the goals and objectives to be achieved, and these shared goals are clear, realistic and feasible, 
\item There is a prevailing notion of democracy and equality among its members, with no one having particular control over the others, 
\item Members perceive a personal development as the team enhances their creativity and provides general support in fulfilling their individual plans.
\end{enumerate}

\paragraph{Analysis.}
Regarding team composition, there is a strong relationship between task type and team type (structure). The type of the team depends on the features of the task to perform and so very often team types are derived from task types.
Besides task type, team diversity plays an important role when composing teams.
Regarding the ``optimal'' team size, it is a complex question and future research is needed to determine the impact of team size on team performance, such as the nature of the task, the internal motivations, and the context. Some preliminary results show that the more complex the task, the larger the size of the team needs to be, but limited to an optimal size of six members. Regarding team formation, several different team processes classifications have been proposed, though no agreement has been reached. Finally, having a good team climate seems key to achieve good performance.

\subsection{WHEN do we do it? The dynamics} \label{whencs}

Humans learn with every interaction. Our memory recollection and capability improvement cannot be removed or stopped. Hence, the organisational psychology research usually deals with complex scenarios, those of simultaneous and successive tasks, see Section \ref{whencs}. In organisational psychology, the dynamic properties of a team are referred to as emergent states. Emergent states develop during teamwork and have an effect on the outcomes. Several examples of emergent states \citep{Mathieuetal} are team confidence, team empowerment, cohesion, team climate, collective cognition or trust between team members.

The development of emergent states is closely connected to the process of team learning behaviours. As members of a team interact with one another and perform tasks, they learn from their experiences. That is, they learn by asking questions, seeking feedback, experimenting, reflecting on results, and discussing errors or unexpected outcomes of previous actions \citep{Edmondson}. These complex tasks allow team members to acquire, share, combine and apply knowledge \citep{ArgoteOlivera,KozlowskiIlgen}. They also lead to the development of shared understanding and meaning as well as to the acquisition of mutual knowledge, skills, and performance capabilities \citep{GaravanMcCarthy}. All these developments enhance team performance \citep{Edmondson,ZellmerBruhnGibson}.

\paragraph{Analysis.} Unlike computer science, the reviewed organisational psychology literature does not study simple scenarios such as non successive and non simultaneous tasks. Typically, organisational psychology analyzes complex and realistic scenarios as human learning capabilities need to be considered. Moreover, on top of including the social network and memory about the outcomes of past experiences, the researchers in organisational psychology deal with the dynamics of individuals' capabilities (as humans learn new capabilities and forget not used ones).

\subsection{WHERE do we do it? The context} \label{whereOP}

From a systemic perspective teams are part of the structure of an organisation and therefore they operate within this organisation. In the same way, an organisation is part of the environment. 
The environment creates demands and requirements for an organisation and influences the organisation's system. In turn, the organisation tries to address these requirements by influencing the operations of its teams and their performance in diverse ways. 

Research results suggest that context plays an important role in the performance of teams \citep{guzzo1996,hackman1990groups}. \cite{hackman1990groups} between others propose and analyse many contextual factors that have to be considered when composing a team:
\begin{itemize}
\item The uncertainty on the level of complexity of the tasks and the degree of dynamics of the environment. Both aspects influence the uncertainty within the organisation and therefore its teams need to operate with incomplete knowledge. The uncertainty about external factors is determined by the available information about the customers, the suppliers, or other competing organisations. The uncertainty about internal factors is determined by the dynamics of tasks, organisational rules and objectives. In such an uncertain context, teamwork is more challenging and paradoxically teams may perform better than in a stable and predictable context.
\item The vision and mission of an organisation that determine the main rules and norms to be followed and what is to be considered as good performance.
\item The set of values, policies and strategies of the organisation. 
For instance, organisations supporting individual values will hinder teamwork and team performance will thus be poor. This is because teamwork is based on shared values, mutual support, constructive collaboration, mutual trust, coordination mechanisms and synergies, which are collective values. On top of it, an organisation promoting internal competition will lead to individual strategies of withholding information and self-interested behaviours.
\item The organisational benefits such as the reward or the training systems. Diverse motivational theories are available to explain the relevance of the reward systems for increased performance. For example, teams will perform better with an appropriate reward system. 
\item The resources and assistance made available to the team. It is obviously easier for the team to achieve good performance when operating in a context of resource abundance.
\item The organisational climate. A context with a perceived climate of control and low level of autonomy for the team will hinder successful teamwork and performance. As teamwork requires an individual engagement with the team, a climate is needed that facilitates information sharing or team skills development. 
\item The cultural context. The definition of a team changes across cultures: in cultures valuing individualism teams are seen more as a set of people each contributing to a different subtask, whereas in cultures valuing collectivism teams are seen as having shared goals, values and responsibility for the whole task. Research results show that teams perform better in a collective cultural context.
\end{itemize}

 \paragraph{Analysis.}
In contrast with computer science approaches, the context where teams solve tasks plays an important role in the organisational psychology literature. The context is understood as internal and external factors influencing teamwork. The internal context can be characterised as dimensions of the organisation, such as vision and mission,  
values, policies and strategies, or organisational benefit system. The external context can be characterized as dimensions of the environment in which the organisation operates, that is the culture, the available resources, and the uncertainty about other players behaviour. 

\section{Discussion}\label{Dis}

Computer Science (CS) and Organisational Psychology (OP) have followed rather disparate approaches when it comes to team composition and team formation. However, some similarities and differences can be drawn and several new research questions can be formulated from a cross reading of the two literature corpus. In Table \ref{tab:CStable} a comparison of the main papers in CS can be found.

\pagestyle{empty}
\begin{landscape}
\begin{table}[htbp]
\begin{center}
\hspace{-2cm}
\begin{tabularx}{26.5cm}{ |X|p{24mm}|p{22mm}|p{22mm}|p{22mm}|p{22mm}|p{23mm}|p{30mm}|p{22mm}|}
    \hline
        {\bf Article Title} & {\bf Team Process}	&  {\bf Individual Properties} & {\bf The task } & {\bf The Objective} 	& {\bf Team Composition}	&	{\bf Team Organisation}	& {\bf The dynamics}  & {\bf The context} \\ 
        \hline
        \cite{Agmon2014} & Formation & Personality & Plan-based & Maximizing social welfare & Exogenous & Hierarchy & Non Successive and Non Simultaneous & N/A
\\ \hline
            \cite{Anagnostopoulos12onlineteam}	& Composition	& Capacity	& Individual-based & Maximizing the quality & Exogenous	& Egalitarian	& Successive and Non Simultaneous & Social and Geographical \\ \hline
        \cite{Andrejczuk} & Composition & Personality & Individual-based & Maximizing the quality & Exogenous & Egalitarian & Non Successive and Simultaneous
& N/A \\ \hline
\cite{Barrett2013} & Formation & Personality & Plan-based & Maximizing the quality & Exogenous & Egalitarian & Non Successive and Non Simultaneous & N/A
\\ \hline
\cite{Chalkiadakis2012} & Composition & Capacity & Individual-based & Maximizing social welfare & Endogenous & Egalitarian & Successive and Simultaneous
& N/A
\\ \hline
\cite{Chen2015} & Composition & Capacity & Individual-based & Maximizing social welfare & Exogenous & Egalitarian & Successive and Simultaneous & N/A
\\ \hline
\cite{Crawford} & Composition & Capacity & Individual-based & Minimizing cost & Exogenous & Egalitarian & Non Successive and Non Simultaneous & N/A
\\ \hline
    \cite{FarhangianPPS15} & Composition & Personality & Individual-based & Maximizing social welfare & Endogenous & Hierarchy & Successive and Simultaneous & N/A \\ \hline
    
    \cite{Farhangian2015} & Composition & Capacity and Personality & Individual-based & Minimizing cost & Exogenous & Egalitarian & Non Successive and Non Simultaneous & N/A\\ \hline
    
    \cite{Hanna2015} & Formation & Personality & Individual-based & Maximizing the quality & Exogenous & Egalitarian & Non Successive and Non Simultaneous
& N/A \\ \hline
    \cite{Kargar} & Composition & Capacity & Individual-based & Minimizing cost & Exogenous & Egalitarian & Non Successive and Non Simultaneous
& N/A \\ \hline
    
    \end{tabularx}
\end{center}
\end{table}\label{tablecorpus}
\end{landscape}

\pagestyle{plain}
\pagestyle{empty}
\begin{landscape}
\begin{table}[htbp]
\begin{center}
\vspace{1cm}\hspace{-2cm}
\begin{tabularx}{26.5cm}{ |X|p{24mm}|p{22mm}|p{22mm}|p{22mm}|p{22mm}|p{23mm}|p{30mm}|p{22mm}|}
    \hline
        {\bf Article Title} & {\bf Team Process}	&  {\bf Individual Properties} & {\bf The task } & {\bf The Objective} 	& {\bf Team Composition}	&	{\bf Team Organisation}	& {\bf The dynamics}  & {\bf The context}\\ 
        \hline

\cite{Liemhetcharat2012} & Composition & Capacity & Individual-based & Maximizing the quality & Exogenous & Egalitarian & Successive and Non Simultaneous & Social
\\ \hline
    \cite{MarcolinoIJCAI2013} & Formation & Personality & Plan-based & Maximizing the quality & Exogenous & Egalitarian & Non Successive and Non Simultaneous
& N/A \\ \hline
\cite{Marcolino2015} & Formation & Personality & Plan-based & Maximizing the quality & Exogenous & Egalitarian & Non Successive and Non Simultaneous & N/A
\\ \hline
\cite{Marcolino2016} & Formation & Personality & Plan-based & Maximizing the quality & Exogenous & Egalitarian & Successive and Non Simultaneous & N/A
\\ \hline
\cite{Okimoto} & Composition & Capacity & Individual-based & Minimizing cost & Exogenous & Egalitarian & Non Successive and Non Simultaneous & N/A
\\ \hline
\cite{JAR2015} & Composition & Capacity & Individual-based & Maximizing the quality & Endogenous & Hierarchy & Successive and Simultaneous
& Social
\\ \hline
    \cite{Rangapuram2015} &	Composition	& Capacity	& Individual-based & Maximizing the quality	&  Exogenous &	Egalitarian / Hierarchy &	Non Successive and Non Simultaneous & Social\\ \hline
   
   \cite{Rochlin} & Formation & N/A & Plan-based & Maximizing the quality & N/A & Hierarchy & Non Successive and Non Simultaneous & N/A
\\ \hline
        \cite{Rokicki}	& Composition and Formation	& N/A & N/A & Maximizing social welfare	& Exogenous, Endogenous	& Egalitarian 	& Non Successive and Simultaneous & N/A\\ 
    \hline
\cite{Spradling2013} & Composition and Formation & Capacity & Individual-based & N/A & Exogenous & Egalitarian & Non Successive and Simultaneous & N/A
\\ \hline

    \end{tabularx}
\end{center}\caption{Comparison of the computer science contributions reviewed in this paper.}
\label{tab:CStable}
\end{table}
\end{landscape}
\pagestyle{plain}

\subsection{Similarities in both approaches}

When modeling agents' properties in CS, there are two main approaches. There is either extensive a-priori information about teammates given as input or ad-hoc scenarios where agents learn their teammates' capabilities. In OP a number of tests are proposed to acquire a-priori information about teammates, such as intelligence or cognitive competences tests, assessment centres or social and behavioural competence tests. Also, similar to CS, OP studies allow to learn human capabilities from their repeated interactions.

To maximize team performance, one of the crucial findings in both OP and CS is that team members have to be heterogeneous.

Regarding the tasks that are executed by agent teams, both OP and CS focus rather on team members' properties required to perform a task than on a detailed planning of the task execution.

\subsection{Differences in both approaches}

The first difference we find between CS and OP is with respect to the complexity of individual team members. Organisational psychology focuses on humans with all their intrinsic complexity while CS  focuses on a limited set of human-like properties to build software agents.
In CS the agent properties have been categorized as personality and capacity. 
In OP, although human properties can also be categorized as personality and capacity, capacity is a much wider concept. It contains not only skills, but also other properties, such as competences, experience, gender or age. Moreover, while in OP the human capabilities are assumed to be dynamic (i.e. lifelong learning), software agents capabilities are assumed to be static and only the behaviour model may change with agents' interactions. 

In CS the majority of approaches assume that the joint capabilities of agents in a team are enough to solve a given task. However, the researchers in OP recognize also other factors as important when composing and forming a team, such as the motivation of individuals and the task context. They also show that the motivation characteristics predict more accurately the performance of a team than the other factors. 
Regarding OP research gaps, it lacks a mapping between cognitive ability of individuals and task types (which is an input in CS models) which complicates team composition.

The CS literature has focused on team co-operation with various objectives that can be categorized as at least one of the following: minimizing overall cost, maximizing social utility, or maximizing the quality of the outcome (understood as maximizing team performance). In OP, the main objective for team composition and formation is just to maximize team performance.
Moreover, from an OP perspective team performance cannot be assessed by the time spent to perform a task, by comparing costs or by counting the number of right answers as it would ignore some important subjective reasons. 
Instead, OP analyzes possible causes of failure, such as an excessive amount of work needed to execute the task given the size of the team or the lack of motivation of team members. This is why the performance is assessed from two perspectives: objective and subjective, while, CS only considers objective measures.
In CS there are only early attempts to include a subjective perspective while analyzing team performance. It is shown that the motivation increases by introducing competition mechanisms (like in crowdsourcing teams) or by giving agents freedom while selecting their collaborators (like in ad-hoc teams).

Since in CS agents can be modeled depending on the needs, researchers can study different settings depending on the dynamics of task arrival (one task or many, one time or many). 
Many MAS models are simplistic since they consider only one task arriving at a time. Unlike CS, the reviewed OP literature does not study simple scenarios, since humans have memory and improve their capabilities with every task. Hence, typically OP analyzes only complex and realistic scenarios. The CS literature uses these complex scenarios to let agents build their beliefs based on past experiences and compose new teams according to these learned beliefs. OP, on top of including the social network and memory about the outcomes of past experiences, deals with the dynamism of individuals' capabilities (as humans learn new capabilities and forget not used ones). 

\subsection{Cross fertilization opportunities}

Although some of the individual properties studied in OP (e.g. age or gender) may not make sense in a CS context, some do. For instance, the dynamics of competences through learning and experience and the cultural values could be used to program more sophisticated agents, specially when interacting in mixed teams involving humans.

The majority of MAS literature on team composition and teamwork assumes that the agents always behave according to their capabilities and knowledge. OP highlights the importance of the motivation of individuals, when estimating performance. There is an opportunity to extend current MAS models by adding agent motivation properties.

Additionally, although in both CS and OP the modeling of the individuals' properties has been broadly studied,  there is still a need in both fields of modeling the properties of agent teams, other than a sum of agents' individual capabilities or a boolean representation of whether the team can perform a task.

In both CS and OP literature, there are some preliminary attempts to include planning, though they are very simplistic. The majority of methods do not consider time constraints, action dependencies, action failure, plan robustness, task requirement dynamic changes and hence, the vast literature on planning has not yet been integrated into team formation methods in both fields.

According to OP having a good team climate seems key to achieve good performance. However, only few CS works recognize team climate (expressed as a synergy or a compatibility graph) as an important factor when composing teams. Further work is needed to investigate the relation between good team climate and team performance in CS research.

In OP, context is considered one of the most important characteristics related to team performance. To our best knowledge, there are only few works in CS that would recognize context as an important factor besides the social and geographical context considered in a few papers. There is a need to perform further research on context to build better performing agent teams.

According to OP there is a strong relationship between task type and team type (structure). However, the majority of CS literature does not correlate team type with a task type apart from the relationship between the number of agents in a team and the set of capacities defined by a task. There is a need to further explore OP task types and their influence on teams' performance.

Finally, despite of a vast body of OP research over decades on team composition, the researchers are not yet at the point of creating the algorithms that lead to the dream team. This survey provides some guidelines on team composition and formation from the CS literature that can help on this lack of formal models and algorithms in OP. 



\section*{Acknowledgements}
Work supported by projects Collectiveware TIN2015-66863-C2-1-R (MINECO/FEDER), SMA (201550E040), and Gencat 2014 SGR 118. The first author is supported by an Industrial PhD scholarship from the Generalitat de Catalunya.

\bibliographystyle{plainnat}

\end{document}